\documentclass{JINST}

\usepackage{graphicx}
\usepackage{multirow}
\usepackage{amssymb,amsmath}

\title{Signal propagation and spark mitigation in resistive strips read-outs.}

\author{Javier Gal\'an$^a$ \\
\llap{$^a$}CEA Saclay,\\
Gif-sur-Yvette, 91191 cedex, France \\
  E-mail: \email{javier.galan.lacarra@cern.ch}}

\abstract{ MicroPattern Gaseous Detectors (MPGD) made of resistive strips have raised as a promising technology for the protection against spark processes having place in the gaseous chamber. The reproduction of the signals and its propagation through the resistive foil is mandatory to better understand its behavior and optimize the key parameters which might depend on the application requirements. In this work it will be presented a resistive-strip model and the charge diffusion through the resistive strip for different model parameters, such as the strip linear resistivity and capacitance, together with the advantages and/or disadvantages of this type of technology. }


\begin{document}

\section{Introduction}

High amplification gains are required in MicroPattern Gaseous Detectors (MPGD) in order to achieve optimum signal to noise ratios, increasing the performance of detectors in terms of energy resolution and efficiency. The gain applied allows to observe signals from gas ionizing interactions which produce few primary electrons at the detector conversion region. The populated electron avalanches achieved with these few primary electrons entail the risk to produce a discharge at the cathode of the detector when the Raether critical electron density is reached, around $\sim 10^7 -  10^8$ electrons per avalanche~\cite{bib7}. Once a particular detector gain has been fixed, different type of interactions inside the chamber (or produced in the chamber structure) can produce a spark with a certain probability which depends on the number of primary electrons generated~\cite{bib6}.

\vspace{0.2cm}

Nowadays, gaseous detector technology is limited by the presence of sparks in hazard and/or high particles flux environments, due to the increased frequency of appearance. Detector discharges might affect the detector response in different ways; \emph{reducing its operating lifetime} due to intense currents produced in short periods of time, heating and melting the materials at the affected regions, \emph{damaging the read-out electronics} which have to support huge current loads in a brief period of time, and moreover \emph{increasing the detector dead-time} given that spark phenomena entail the \emph{discharge of the cathode} and therefore the amplification field is lost during a relatively long period of time, which is required by the high voltage power supply to restore the charges and recover the nominal voltage.

\vspace{0.2cm}

It was first observed in RPC-type\footnote{Resistive Plate Chambers} detectors that the introduction of a high impedance resistive coating at the anode limits the detector current during a period of at least some $\mu$s, constraining the spark process to the streamer phase and reducing the total amount of charge released~\cite{bib0}. Furthermore, the limited discharge current affects only locally avoiding the field to be lost at the remaining detector regions, and thus reducing the dead-time of the detector.

\vspace{0.2cm}

Micromegas detectors were introduced already in 1995~\cite{bib1} as a good candidate for high particle flux environments, and spark studies with standard micromegas technology were also carried out~\cite{bib3}. Recently, additional efforts are pushing the development of resistive strip micromegas detectors in order to increase its robustness and protection in hazard or high flux environments by limiting spark discharges in the same way as it was done for RPCs. In particular, the MAMMA\footnote{Muon ATLAS MicroMegas Activity} collaboration is developing large area micromegas detectors, which introduce the resistive coating technique~\cite{bib8}, for the upcoming upgrade of the HL-LHC\footnote{High Luminosity Large Hadron Collider}, where the luminosity will be increased by a factor 10 reaching up to $L = 10^{35}$cm$^{-2}$s$^{-1}$.

\vspace{0.2cm}

The MAMMA collaboration has investigated new detector prototypes, with different resistive coating topologies. These studies have increased the robustness and stability of micromegas detectors in the presence of an intense and highly ionizing environment, limiting the negative effect of sparks could have on them~\cite{bib5}.

\vspace{0.2cm}

The present work pretends to give a theoretical description of the signal propagation through a resistive strip as it was previously done in RPC-type detectors~\cite{bib2}. A resistive strip model will be introduced based on the resistive-strip detectors developed in the framework of the MAMMA collaboration. The intention of this work is to provide a model of the electric behavior of the detector to a current signal in order to better understand the different responses of the detector which could be produced in different environments, and moreover the response dependency on the key parameters of the detector, which will be defined in the model. It is expected, that the future agreement between model and experimental measurements could allow to exploit the still, if any, unexplored capabilities of this new type of technology.







\section{ The resistive strip model.}

A simplified approach of a single resistive strip is taken into to account in order to calculate the induced signal at the micromegas usual read-out and the charge diffusion through the resistive strip. The model here described is based on the equivalent electric circuit introduced at~\cite{bib5}. This equivalent circuit describes in general a standard micromegas detector covered by a thin insulating layer where resistive strips are placed facing the standard metallic strips, creating an inter-capacitance and therefore an electric transmission line.

\vspace{0.2cm}

\begin{figure}[!ht]\begin{center}
\includegraphics[width=0.9\textwidth]{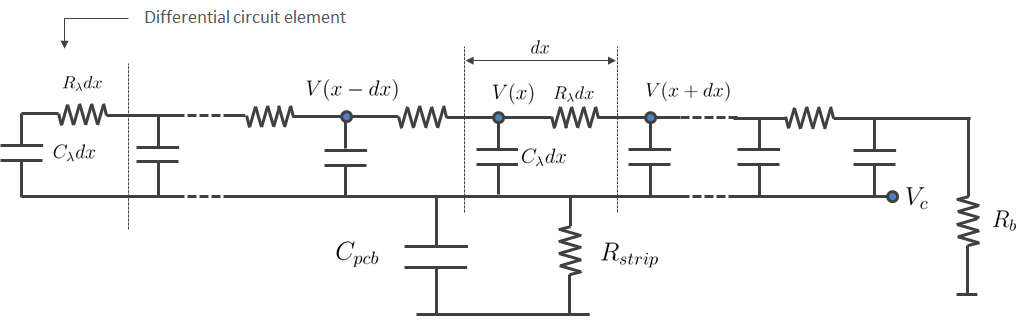}
\caption{\fontfamily{ptm}\selectfont{\normalsize{ Simplified RC-model of a resistive strip.   }}}
\label{resistive_RCmodel}
\end{center}\end{figure}

In order to calculate the charge diffusion, the resistive strip is subdivided in a set of differential elements which final element is grounded through a boundary resistor $R_b$. The resistive strip has a linear resistivity $R_\lambda$ and is electrically connected to the standard read-out, defined by $V_c$, through the mentioned insulating layer which introduces the strip linear capacitance $C_\lambda$. The standard read-out is connected in this model to a low impedance value $R_{strip}$, and the residual capacitance of the PCB board  $C_{pcb}$ (see Fig.~\ref{resistive_RCmodel}). 

\vspace{0.2cm}

In this simplified model, applying Kirchoff continuity at each node we can describe the resistive strip potential $V(x,t)$ in terms of the current induced at the inter-capacitance $C_{\lambda}$

\begin{equation}
\frac{V(x+\delta x,t) - V(x,t)}{R_\lambda \delta x} - \frac{V(x,t)-V(x-\delta x,t)}{R_\lambda \delta x} = C_\lambda \delta x \left( \frac{\partial V(x,t)}{\partial t} - \frac{dV_c(t)}{dt} \right)
\end{equation}

\vspace{0.2cm}
\noindent in the case where $\delta x \rightarrow 0$ the above equation is reduced to the following expression

\vspace{0.2cm}

\begin{equation}\label{eq:diffusion_1}
\frac{\partial^2 V(x,t)}{\partial x^2} = C_\lambda R_\lambda \frac{\partial\left( V(x,t) - V_c(t) \right) }{\partial t }
\end{equation}

\vspace{0.2cm}

\noindent a partial differential equation (PDE= which describes the spatial and temporal evolution of the potential in the resistive strip, and thus the charge distribution in the strip which is directly related with the potential $V(x,t)$ by the relation

\begin{equation}
Q_\lambda(x,t) = C_\lambda \left( V(x,t) - V_c(t) \right)
\end{equation}

\vspace{0.2cm}
\noindent where $Q_\lambda$ is the linear charge density in the resistive strip.

\vspace{0.2cm}
The evolution of the potential and/or charge in the resistive strip is bounded by the external circuit elements attached to the strip electrode, $R_{strip}$ and $C_{pcb}$, and the fixed resistive value, $R_b$, attached to the resistive strip. The first elements ($R_{strip}$ and $C_{pcb}$) allow to determine the strip voltage $V_c$ as a function of the resistive strip potential distribution, by integrating the total equivalent current induced at the capacitive surface $i_{C_\lambda}$ along the resistive strip length $x_L$,

\begin{equation}
i_{C_\lambda} = C_\lambda \int_0^{x_L} \frac{\partial}{\partial t} ( V(x,t)-V_c(t) ) dx = C_{pcb} \frac{dV_c(t)}{dt} + \frac{V_c(t)}{R_{strip}}
\end{equation}

\noindent relation that can be rewritten in the most convenient expression

\begin{equation}\label{eq:dVcdt}
\frac{dV_c(t)}{dt} = \frac{ C_\lambda } {C_{pcb} + x_L C_\lambda} \int_0^{x_L} \frac{\partial V(x,t)}{\partial t}  dx - \frac{V_c(t)}{ (x_L C_\lambda + C_{pcb})  R_{strip}}
\end{equation}

\vspace{0.2cm}

The resistive value $R_b$ provides the boundary condition on the voltage $V_o(t)$ at the beginning of the strip, and its value is related with the voltage gradient at the resistive strip end,

\begin{equation}
V_o(t) = -\frac{R_b}{R_\lambda} \frac{\partial V(x,t)}{\partial x} \Big |_{x=x_o}.
\end{equation}

\vspace{0.2cm}

The relation (\ref{eq:diffusion_1}) describes the potential propagation at the resistive strip. However, in order to observe such propagation we need a perturbation which is introduced as a time dependent current source spatially distributed and described by the change on linear charge density $\rho(x,t)$ along the resistive strip. This current is introduced inside relation (\ref{eq:diffusion_1}) by taking into account the expected voltage increase at the resistive nodes,

\begin{equation}\label{eq:diffusion_2}
\frac{\partial^2 V(x,t)}{\partial x^2} = C_\lambda R_\lambda \frac{\partial\left( V(x,t) - V_c(t) \right) }{\partial t } + R_\lambda \frac{\partial \rho (x,t)}{\partial t}
\end{equation}

\vspace{0.2cm}
\noindent the introduced charge density pretends to emulate the effect of charge induced by a physical avalanche, which would be the connection between our detector read-out and the physical signal.

\vspace{0.2cm}

The solution of the coupled relations (\ref{eq:dVcdt}) and (\ref{eq:diffusion_2}) for a given input charge density $\rho(x,t)$ will provide a full description of the signal propagation within the frame of this model.

\section{Discretization and resistive strip potential solution.}

If the resistive strip is subdivided in a total of $N$ differential elements of length $\delta x$, the discrete model contains a total of $N+1$ resistive potential nodes. The discrete relations of the $N-1$ internal nodes is obtained directly from the relation~(\ref{eq:diffusion_2}), describing the potential evolution in terms of the neighborhood potentials and the equivalent current at the given node,

\begin{equation}\label{eq:discretization_1}
\frac{dv_j}{dt} = \frac{1}{\tau_\lambda \delta x^2} \left( v_{j+1} - 2 v_j + v_{j-1} \right) + \frac{dv_c}{dt} - \frac{1}{C_\lambda} \frac{d\rho_j}{dt}
\end{equation}

\vspace{0.2cm}
\noindent where $\tau_\lambda = R_\lambda C_\lambda$. The potential at the last node, which defines the non-grounded strip end is given by the relation

\begin{equation}\label{eq:discretization_2}
\frac{dv_N}{dt} = - \frac{1}{\tau_\lambda \delta x^2} \left( - v_N + v_{N-1}\right) + \frac{dv_c}{dt} - \frac{1}{C_\lambda}\frac{d\rho_N}{dt}
\end{equation}

\vspace{0.2cm}
\noindent and the first strip element which is constrained with the resistor $R_b$ leads to the following boundary relation

\begin{equation}\label{eq:boundary}
\frac{dv_o}{dt} = \frac{1}{\tau_\lambda \delta x^2} \left[ -\left( 1 + \frac{\delta x R_\lambda}{R_b} \right) v_o + v_1\right] + \frac{dv_c}{dt} - \frac{1}{C_\lambda}\frac{d\rho_o}{dt}.
\end{equation}

\vspace{0.2cm}
This set of $N+1$ equations can be written in the matricial form

\begin{equation}
\frac{d}{dt}
\begin{bmatrix} 
v_0	\\
\vdots	\\
v_N	\\
\end{bmatrix}
=
\frac{1}{\tau_\lambda \delta x^2}
\begin{bmatrix}
-\left(1 + \frac{R_\lambda \delta_x}{R_b}\right)	&	1	&		&		& 	0	\\
1	&	-2	&	1	&		& 	0	\\
	&	\ddots	&	\ddots	&	\ddots	&		\\
	&		&	1	&	-2	&	1	\\
0	&		&		&	1	&	-1	\\
\end{bmatrix}
\begin{bmatrix} 
v_0	\\
v_1	\\
\vdots	\\
v_{N-1}	\\
v_N	\\
\end{bmatrix}
-\frac{1}{C_\lambda}
\frac{d}{dt}
\begin{bmatrix} 
\rho_0 	\\
\vdots	\\
\rho_N 	\\
\end{bmatrix}
+\frac{dV_c}{dt}
\begin{bmatrix} 
1 	\\
\vdots	\\
1 	\\
\end{bmatrix}
\end{equation}
\noindent which can be reduced to the following expression that represents a set of $N+1$ coupled differential equations

\begin{equation}\label{eq:firstDiscrete}
\frac{d\mathbf{v}}{dt} = \frac{1}{\tau_\lambda \delta x^2} \mathcal{P} \, \mathbf{v} - \frac{1}{C_\lambda} \frac{d\mathbf{\rho}}{dt} + \frac{d V_c }{dt} \, \mathbf{b}.
\end{equation}

\vspace{0.2cm}
Furthermore, the potential $V_c$ at the strip read-out, which is described by the relation~(\ref{eq:dVcdt}), depends on the resistive potential $V(x,t)$. In order to introduce this dependency inside relation~(\ref{eq:firstDiscrete}), the relation describing $dV_c/dt$ must be discretized as follows,

\begin{equation}\label{eq:discretization_3}
\frac{dV_c}{dt} = - \frac{\xi V_c}{ C_\lambda R_{strip}} + \xi \delta x \frac{d}{dt} \sum_0^n v_j
\end{equation}

\vspace{0.2cm}
\noindent where $\xi = C_\lambda/(C_{pcb} + x_L C_\lambda)$. Introducing the last expression inside relation~(\ref{eq:firstDiscrete}) we obtain

\begin{equation}\label{eq:secondDiscrete}
\mathcal{H}\frac{d\mathbf{v}}{dt} = \frac{1}{\tau_\lambda \delta x^2} \mathcal{P} \, \mathbf{v} +  \frac{1}{\tau_\lambda \delta x^2} \, \mathbf{v_o} - \frac{1}{C_\lambda} \frac{d\mathbf{\rho}}{dt} -\frac{ \xi V_c }{C_\lambda R_{strip}} \, \mathbf{b}
\end{equation}

\vspace{0.2cm}
\noindent where the matrix~$\mathcal{H}$ is defined as 

\begin{equation}
\mathcal{H}=
\begin{bmatrix}
1 - \xi \delta x	&		& 	-\xi \delta x	\\
			&	\ddots	&			\\
-\xi \delta x		&		&	1 - \xi \delta x	\\
\end{bmatrix}.
\end{equation}

\vspace{0.2cm}
Finally, the differential dependency on $V_c$ has been substituted by a linear dependency on $V_c$, which can be more easily introduced in the calculation by imposing some initial conditions. If we work out the value of $d\mathbf{v}/dt$ we obtain the following final expression for the potential change along the resistive strip

\begin{equation}\label{eq:finalDiscrete}
\frac{d\mathbf{v}}{dt} = \frac{1}{\tau_\lambda \delta x^2} \mathcal{H}^{-1} \mathcal{P} \, \mathbf{v} +  \frac{1}{\tau_\lambda \delta x^2} \, \mathcal{H}^{-1} \mathbf{v_o} - \frac{1}{C_\lambda} \mathcal{H}^{-1} \frac{d\mathbf{\rho}}{dt} -\frac{ \xi V_c }{C_\lambda R_{strip}}  \mathcal{H}^{-1} \, \mathbf{b}
\end{equation}

\vspace{0.2cm}
\noindent expression which represents a set of $N+1$ coupled first order differential equations. In order to solve the potential it is required to uncouple this set of equations and apply the standard method to solve a set of independent first order differential equations. To achieve that the matrix $\mathcal{D}$ which diagonalizes $\mathcal{H}^{-1}\mathcal{P}$ is introduced in the calculation

\begin{equation}
\Lambda = \mathcal{D}^{-1} \left( \mathcal{H}^{-1} \mathcal{P} \right) \mathcal{D}
\end{equation}


\vspace{0.2cm}
\noindent where $\Lambda$ is the resulting diagonal matrix. By multiplying to the left side the relation~(\ref{eq:finalDiscrete}) by $\mathcal{D}^{-1}$ and introducing the identity $\mathcal{D}\mathcal{D}^{-1}$ the term which contains $\mathbf{v}$ becomes diagonal

\begin{equation}
\begin{split}
\frac{d}{dt} \left( \mathcal{D}^{-1} \mathbf{v} \right) =& \frac{1}{\tau_\lambda \delta x^2} \left[ \mathcal{D}^{-1} \left( \mathcal{H}^{-1} \mathcal{P} \right) \mathcal{D} \right] \left( \mathcal{D}^{-1} \, \mathbf{v} \right)   \\
& - \frac{1}{C_\lambda} \mathcal{D}^{-1} \mathcal{H}^{-1} \frac{d\mathbf{\rho}}{dt} -\frac{ \xi V_c }{C_\lambda R_{strip}}   \mathcal{D}^{-1}\mathcal{H}^{-1} \, \mathbf{b}
\end{split}
\end{equation}

\vspace{0.2cm}
\noindent obtaining a set of differential equations which are uncoupled and where the transformed potential $\mathbf{u} = \mathcal{D}^{-1} \mathbf{v}$ can be solved for each virtual node. Finally, the set of $N+1$ uncoupled differential equations can be better described by the following expression

\begin{equation}\label{eq:finalfinal}
\frac{d\mathbf{u}}{dt} = \frac{1}{\tau_\lambda \delta x^2} \Lambda \mathbf{u} +  \frac{1}{\tau_\lambda \delta x^2} \, \mathcal{X} \mathbf{v_o}  - \frac{1}{C_\lambda} \mathcal{X} \frac{d\mathbf{\rho}}{dt} -\frac{ \xi V_c }{C_\lambda R_{strip}}   \mathcal{X} \, \mathbf{b}
\end{equation}

\vspace{0.2cm}
\noindent where $\mathcal{X} = \mathcal{D}^{-1} \mathcal{H}^{-1}$.

\vspace{0.2cm}
The expression~(\ref{eq:finalfinal}) can be numerically solved for a given time step $\Delta t$ by any of the popular methods (\emph{Euler} or \emph{Runge-Kutta}). The time discretization is now trivial in this relation and the value of $V_c$ can be updated at each time iteration by relation~(\ref{eq:discretization_3}) by calculating first the transformed potentials $\mathbf{u}$ and obtaining the real resistive strip potential $\mathbf{v} = \mathcal{D} \mathbf{u}$ at each time iteration, imposing the initial conditions to satisfy $V_c(t=0) = 0$ and $V(x,t = 0) = 0$.

\section{Charge diffusion through the resistive strip.}

The model described, together with the method to calculate the propagation, has been used to calculate the charge diffusion along the resistive strip. In order to study such propagation a linear current density must be introduced. The signal is chosen to be Gaussian shaped for simplicity. Therefore, at each node $j$, the linear current density can be described by the following expression

\begin{equation}\label{eq:inputCurrent}
\frac{d\rho_j}{dt} = \frac{\lambda_q \delta x}{2\pi \sigma_x\sigma_t} \mbox{exp}\left(-\frac{(x_j-x_m)^2}{2\sigma_x^2}\right) \mbox{exp}\left(-\frac{(t-t_m)^2}{2\sigma_t^2}\right)
\end{equation}

\noindent where $\lambda_q$ describes the central linear charge density of the virtual event, $x_m$ the mean event position, $t_m$ the time at which the maximum current is reached and, $\sigma_x$ and $\sigma_t$ the spatial and temporal charge diffusion previously produced in the gaseous medium.

\vspace{0.2cm}

\begin{figure}[!ht]\begin{center}
\begin{tabular}{cc}
\includegraphics[width=0.475\textwidth]{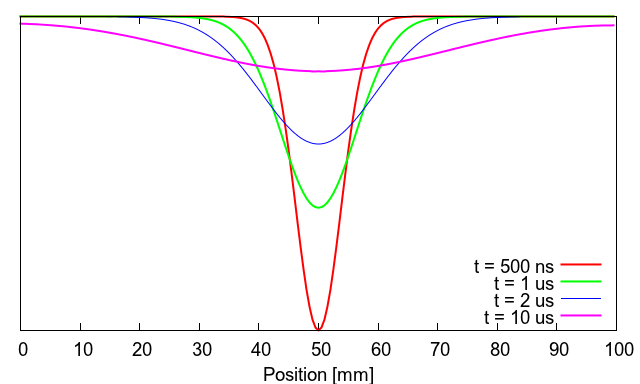} &
\includegraphics[width=0.445\textwidth]{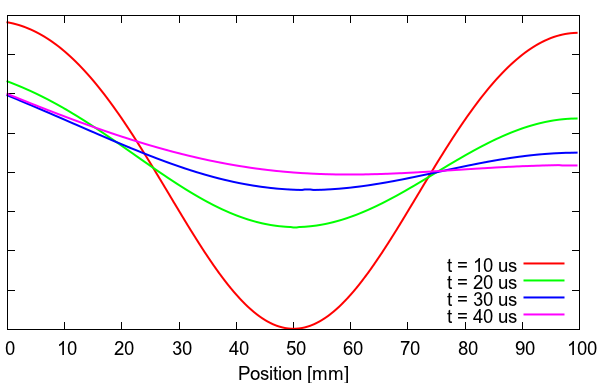} \\
\end{tabular}
\caption{\fontfamily{ptm}\selectfont{\normalsize{ Potential distribution along the resistive strip at different periods of time. On the left, first periods of time, on the right plot, snapshots after longer time period has passed. }}}
\label{chargeDiffVsTime}
\end{center}\end{figure}

A relatively fast signal is generated at $t_m = 200$\,ns with $\sigma_t = 50$\,ns at the central position, $x_m = 5$\,cm with $\sigma_x = 0.5$\,cm. Figure~\ref{chargeDiffVsTime} shows the result of charge diffusion due to such current signal through the resistive strip, $10$\,cm long, at different periods of time.

This first calculation allows to observe the charge spread along the resistive strip as a function of time, and also the asymmetry effect at the boundary resistor which can be observed at the last time steps, which is produced by the electrons drifting towards the ground potential.

\begin{figure}[!ht]\begin{center}
\begin{tabular}{cc}
\includegraphics[width=0.45\textwidth]{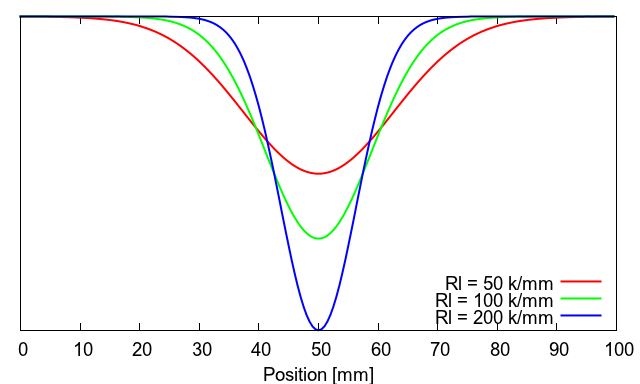} &
\includegraphics[width=0.45\textwidth]{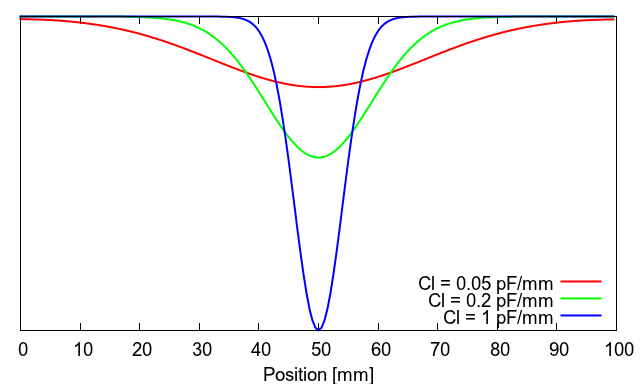} \\
\end{tabular}
\caption{\fontfamily{ptm}\selectfont{\normalsize{ On the left, charge distribution along the resistive strip for different resistivity values at the same period of time. On the right, the distribution for different values of the linear capacitance.  }}}
\label{chargeDiffVsParameters}
\end{center}\end{figure}

\vspace{0.2cm}
The same calculation was performed for different parameter combinations, including linear resistivity $R_\lambda$ and linear capacitance $C_\lambda$. Figure~\ref{chargeDiffVsParameters} shows the effect of those \emph{two} parameters on the charge diffusion, showing a faster charge spread as the resistivity $R_\lambda$ and/or the capacitance $C_\lambda$ are lowered.

\vspace{-0.4cm}
\subsection{Homogeneous current effect on the resistive strip.}

The longer times required for charge diffusion through the resistive strip, and its obvious dependency with the charge position through the time constant $\tau_\lambda [$s$\cdot$cm$^{-2}]$, will affect the potential homogeneity at the resistive anode in high rate and thus intense radiation environments.

\vspace{0.2cm}

\begin{figure}[!ht]\begin{center}
\begin{tabular}{cc}
\includegraphics[width=0.47\textwidth]{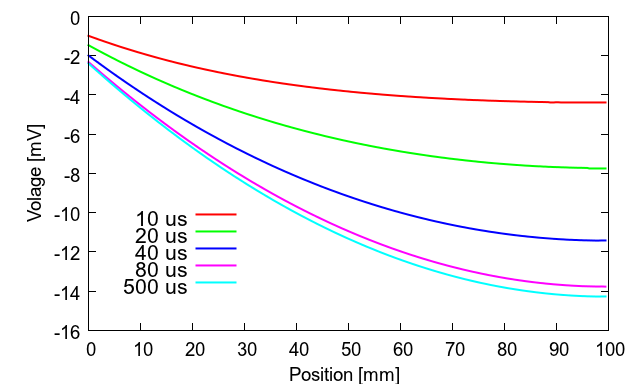} &
\includegraphics[width=0.47\textwidth]{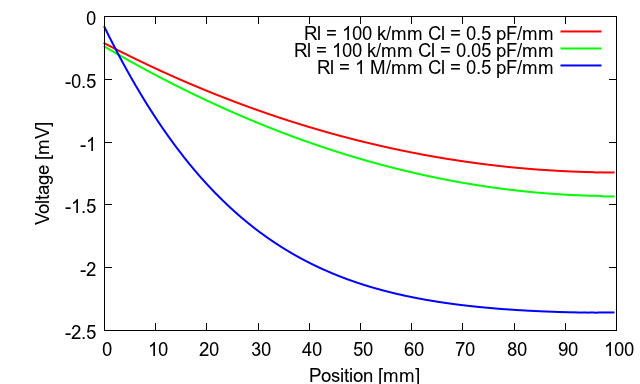} \\
\end{tabular}
\caption{\fontfamily{ptm}\selectfont{\normalsize{On the left, transition effect on potential distribution due to the homogeneous illumination switched on at the starting reference time. On the right, final-state potential distribution for different model parameters.  }}}
\label{constantRateEffect}
\end{center}\end{figure}

The resistive-strip model previously described has been used to show the effect of an homogeneous illumination on the resistive potential distribution $V(x,t)$. This is achieved by imposing the input signal previously given by relation~\ref{eq:inputCurrent} to be uniform and constant in time. Figure~\ref{constantRateEffect} shows the potential $V(x,t)$ transition along the resistive-strip and the effect of the same input current as a function of $R_\lambda$ and $C_\lambda$ on the final state, emphasizing the higher effect with the resistivity.

\subsection{Resistive charge diffusion dependence on event position.}

\begin{figure}[!ht]\begin{center}
\begin{tabular}{cc}
\includegraphics[width=0.46\textwidth]{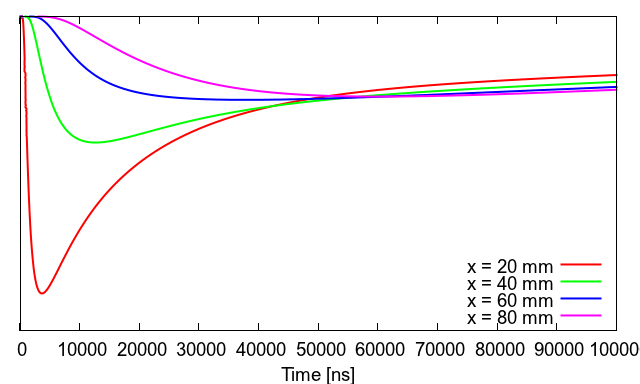} &
\includegraphics[width=0.46\textwidth]{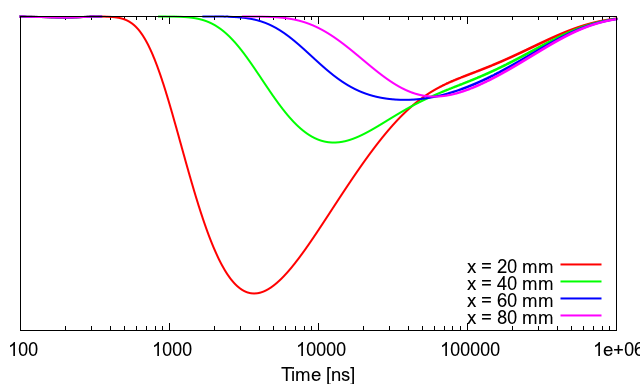} \\
\end{tabular}
\caption{\fontfamily{ptm}\selectfont{\normalsize{ Current at the resistive-strip boundary resistor as a function of time, for events produced at different positions. On the left plot, linear scale and on the right plot, logarithmic scale.  }}}
\label{pulseShapes}
\end{center}\end{figure}

The charge diffusion through the resistive-strip becomes necessarily an effective current at the boundary resistor, coming from charges in the resistive material that drift towards the ground. This current can be easily obtained from the propagation calculated in the model by representing the value of the potential $v_o(t)$ at the boundary resistor as a function of time. The pulse-shape of the signals propagating through the resistive-strip will depend on the event position, an effect that will not be negligible thanks to the time propagation delay induced by the high strip resistivity. Figure~\ref{pulseShapes} shows the current-shape at the boundary resistor for similar characteristics events having place at different strip coordinates, being the lower position values closer to the strip boundary resistor. As it is observed in these plots, the closer the event is to the ground the faster is the expected signal at the boundary resistor.

\begin{figure}[!ht]\begin{center}
\includegraphics[width=0.57\textwidth]{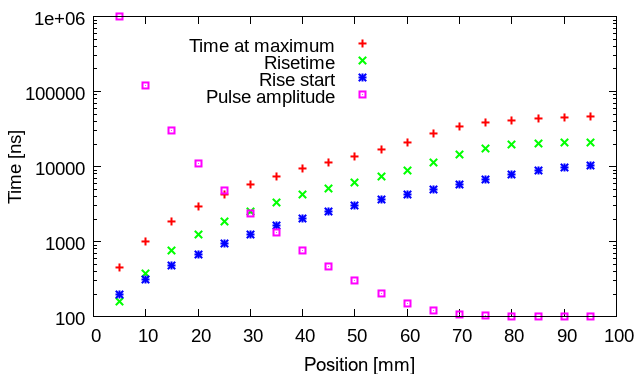}
\caption{\fontfamily{ptm}\selectfont{\normalsize{ Pulse time properties, for a resistivity $R_\lambda = 100$\,k$\Omega/$mm, as a function of the position. The relative pulse amplitude is also represented in an auxiliary linear scale.  }}}
\label{pulsepropertiesWithResisitivity}
\end{center}\end{figure}

\vspace{-0.2cm}
Once the event signal is fixed, the different pulse-shapes can be associated uniquely to a given position. This property of resistive strips has already been used for position detection in silicon based detectors by using the charge division method~\cite{bib4}, introducing an output signal at each end. In our particular case, with one unique end, the position sensitivity could be reached by analyzing the pulse time properties (see Fig.~\ref{pulsepropertiesWithResisitivity}).

The signal properties can be therefore used to localize the position of the event in the resistive strip. It is remarkable how the pulse amplitude is weakened as the event is farther away from the strip-end where the resistor ground is located. This fact will favor the localization of events which are near the ground. But at least, events producing huge amount of charge, as developing streamers, will produce signals which could still be observed.

\vspace{0.2cm}

\begin{figure}[!ht]\begin{center}
\begin{tabular}{cc}
\includegraphics[width=0.47\textwidth]{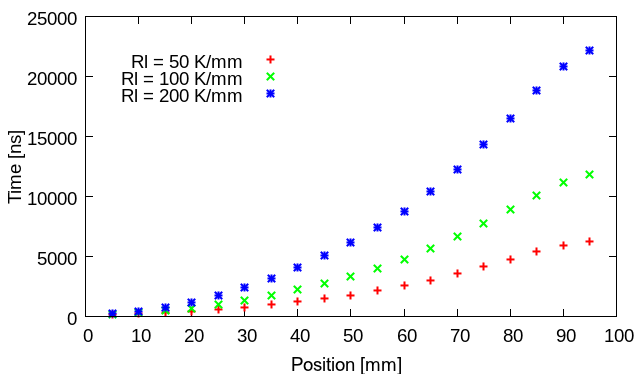} &
\includegraphics[width=0.47\textwidth]{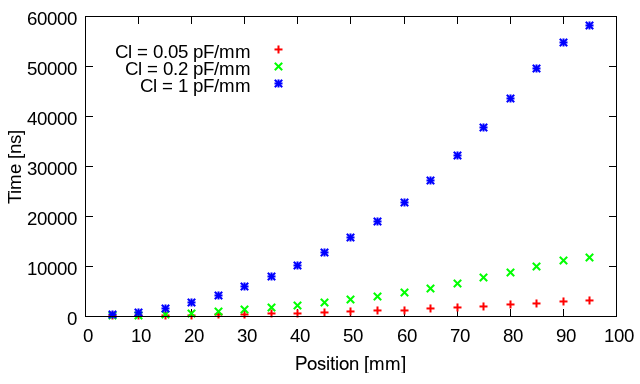} \\
\multicolumn{2}{c}{\includegraphics[width=0.94\textwidth]{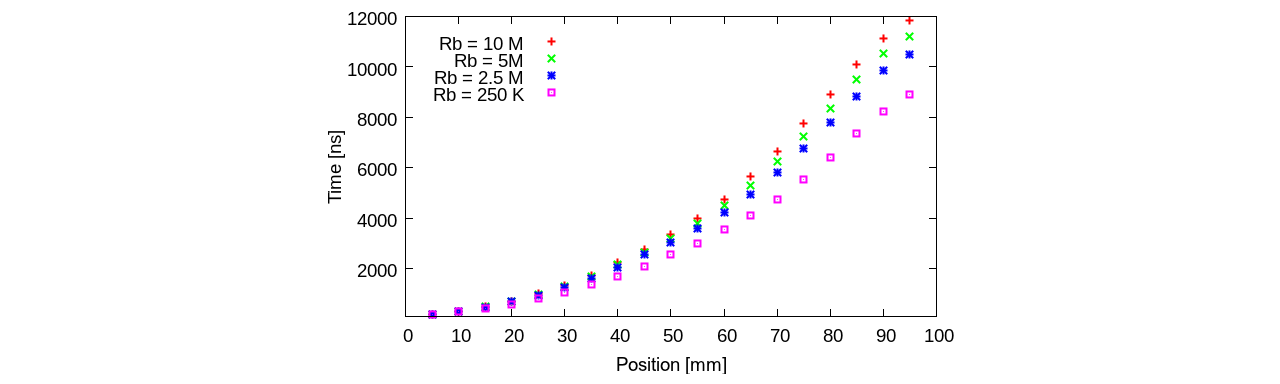}} \\
\end{tabular}
\caption{\fontfamily{ptm}\selectfont{\normalsize{Time signal delay as a function of the event position comparing the effect of different model parameter values.  }}}
\label{pulsePropertiesWithParameters}
\end{center}\end{figure}

The temporal pulse shape properties depend on the resistive parameters such as resistivity $R_\lambda$, capacitance $C_\lambda$ and boundary resistor values $R_b$. Figure~\ref{pulsePropertiesWithParameters} shows the time delay dependency with these parameters as a function of the event position. It is observed how the time delay changes are increased with higher resistivities and capacitances, and the effect of the boundary resistor value is not affecting considerably. A higher change on time delay with position should provide a higher equivalent spatial definition, however a compromise between signal delay and signal intensity should also be considered.

%
\section{Conclusions and future work}

A method for the description of the slow diffusion signal on resistive strip has been formulated. Some results concerning relevant matters of the detector have been presented showing the still predicting potential of a simplified model. The full mathematical description provided for this model could allow future extensions to be implemented, as it could be the case considering resistive-strip inhomogeneities, and other complexities that cannot be addressed with commercial or dedicated simulation software. Moreover, it could provide the key to connect streamer physics and other related physical phenomena with this kind of detectors, allowing to study the response of the detector in a variety of physics scenarios and applications.

\vspace{0.2cm}

\begin{figure}[!ht]\begin{center}
\includegraphics[width=0.9\textwidth]{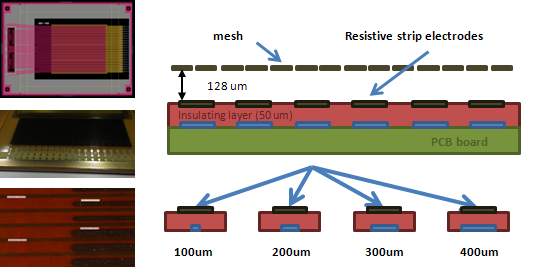}
\caption{\fontfamily{ptm}\selectfont{\normalsize{ On the left, from top to bottom, detector Gerber design,  a picture of resistive strip connectors on detector PCB board, and a microscope picture of the resistive strips with different widths on top of standard strips. On the right, schematic showing the transversal section of the detector.    }}}
\label{prototypes}
\end{center}\end{figure}

New prototypes, based on the MAMMA project ones, have been recently produced by R. Oliveira's team at CERN Micromegas workshop. Three different prototypes have been designed which combine different resistive strip and metallic strip widths in order to provide a single detector with different resistive and capacitive properties. A special contact has been inserted in these prototypes in order to have access to the signal coming out from the resistive strips and at the same time allowing to change the value of the boundary resistor (see Fig.~\ref{prototypes}).

A dedicated electronic read-out is being designed for these prototypes in order to preamplify and integrate the resistive signals with the adequate timing. The results obtained could allow to verify the resistive-strip model described and/or provide clues to apply any modifications required. The measurement of resistive propagation signals in different conditions will be studied for the first time in this type of detectors. Furthermore, the possibility to read-out the position by using this technique could provide a way to reduce the number of read-out channels maintaining good spatial detection resolution.



\begin{thebibliography}{9}

\bibitem{bib7}
H. Raether, 
\emph{Electron avalanches and breakdowns in gases},
Butterworths, Washington, 1964.

\bibitem{bib6}
S. Procureur {\em et al.},
\emph{A Geant4-based study on the origin of the sparks in a Micromegas detector and estimate of the spark probability with hadron beams},
\href{http://dx.doi.org/10.1016/j.nima.2010.05.024}
{\emph{Nucl. Instr. and Meth. A}, {\bf 621} (2010) p. 177}.

\bibitem{bib0}
P. Fonte {\em et al},
\emph{ A spark-protected high-rate detector},
\href{http://dx.doi.org/10.1016/S0168-9002(99)00221-1}
{\emph{Nucl. Instr. and Meth. A}, {\bf 431} (1999) p. 154}.

\bibitem{bib1}
Y. Giomataris {\em et al.},
\emph{MICROMEGAS: A High granularity position sensitive gaseous detector for high particle flux environments},
\href{http://dx.doi.org/10.1016/0168-9002(96)00175-1}
{\emph{Nucl. Instr. and Meth. A}, {\bf 376} (1996) p. 29}.

\bibitem{bib3}
A. Bay {\em et al.},
\emph{Study of sparking in Micromegas chambers},
\href{http://dx.doi.org/10.1016/S0168-9002(02)00510-7}
{\emph{Nucl.\,\,Instr.\,\,and\,\,Meth.\,\,A}, {\bf 488} (2002) p. 162}.

\bibitem{bib8}
T. Alexopoulos {\em et al.},
\emph{Development of large size Micromegas detector for the upgrade of the ATLAS Muon system},
\href{http://dx.doi.org/10.1016/j.nima.2009.06.113}
{\emph{Nucl. Instr. and Meth. A}, {\bf 617} (2010) p. 161}.


\bibitem{bib5}
J. Burnens {\em et al.}, 
\emph{A spark-resistant bulk-micromegas chamber for high-rate applications},
\href{http://dx.doi.org/10.1016/j.nima.2011.03.025}
{\emph{Nucl.\,\,Instr.\,\,and\,\,Meth.\,\,A}, {\bf 640} (2011) p. 110}.

%
\bibitem{bib2}
M. S. Dixit, A. Rankin,
\emph{Simulating the charge dispersion phenomena in micro pattern gas detectors with a resistive anode},
\href{http://dx.doi.org/10.1016/j.nima.2006.06.050}
{\emph{Nucl.\,\,Instr.\,\,and\,\,Meth.\,\,A}, {\bf 566} (2006) p. 281}.

\bibitem{bib4}
J. K. Carman {\em et al.},
\emph{Longitudinal resistive charge division in multi-channel silicon strip sensors},
\href{http://dx.doi.org/10.1016/j.nima.2011.05.016}
{\emph{Nucl.\,\,Instr.\,\,and\,\,Meth.\,\,A}, {\bf 646} (2011) p. 118}.

\end{thebibliography}
\end{document}